\documentstyle[12pt]{article}
\title{{\hfill\normalsize ITP-95-23E}\\[1.0cm]
On the normal phase of 2D Fermi liquid with weak attraction
between particles}

\author{{\sl V.P. Gusynin, V.M.Loktev\thanks{E-mail: vloktev@gluk.apc.org}
and I.A. Shovkovy}\\
{\sl
N.N.Bogolyubov Institute for Theoretical Physics}\\
{\sl Kiev, 252143 Ukraine}}
\date{}

\newcommand{\siml}{\stackrel{\scriptscriptstyle <}{\scriptscriptstyle\sim}}
\newcommand{\xq}{\begin{equation}}
\newcommand{\zq}{\end{equation}}
\newcommand{\beq}{\begin{eqnarray}}
\newcommand{\eeq}{\end{eqnarray}}

\begin{document}
\maketitle

\vfill

\begin{abstract}
Proceeding from the simplest field theoretical model of  2D metal,
the normal phase Green functions of the weakly interacting fermions and
the order parameter fluctuations (responsible for the attraction between
fermions) are obtained. It is shown that taking into consideration the
fluctuations mentioned leads to a considerable reduction of the
fermion wave function renormalization constant (quasiparticle weight)
as well as to a linear dependence of the quasiparticle damping on the
temperature. A general dependence of 2D Fermi liquid properties on the
fermion density is discussed. The relevance of the proposed model to the
marginal behavior of the Fermi liquid of high--$T_c$ superconductors,
in particular, to their linear
temperature dependence of the resistivity is indicated.
\end{abstract}

\vfill
\eject

\section{Introduction}

Theoretical study of the normal properties of 2D (as well as quasi--2D)
metals was strongly intensified by the discovery of high--$T_c$
superconductors (HTSCs). These compounds display a number of nonconventional
features that make it difficult to use the Fermi liquid  (FL) model to
describe their main properties. The most striking features are:
(i) an essentially reduced quasiparticle weight (QW), $Z$, of the excitations
in the vicinity of the Fermi surface; (ii) a rather large damping of the
corresponding quasiparticles; (iii) a linear dependence of the resistivity
on the temperature above $T_c$, and many others (see, for example
\cite{Phys}). Now, as far as we know, there is no satisfactory understanding
as to what makes these compounds behave so strangely\footnote{Sometimes they
are called the "strange metals" \cite{Phys}.} and why they cannot be
described by the theory of Landau FL. Thus, searching for theoretical
models, being able to explain at least some of those odd properties of HTSCs,
remains a very important problem. In this paper we shall make an attempt to
show that one of the possible answers is that the observable properties of
HTSCs can be the result of the extreme susceptibility of 2D attractive
interaction FL with respect to a deep intrinsic structural rearrangement.

We recall that the most characteristic property of the ideal Fermi gas is
the existence of the Fermi energy, $\epsilon_F$, which,  in accordance
with the Pauli principle, separates occupied and empty states. The distribution
of
particles over the energy (and momenta) at $T=0$ can be represented by the
step--function. Adding to the system a weak interaction between the particles
corresponds to the transition from the ideal Fermi gas to the FL. When the
interaction is weak and there is no danger of any kind of instability, the
distribution
function changes only a little bit: the QW slightly decreases ($Z\siml 1$)
and a weak damping, $\gamma\sim (\epsilon-\epsilon_F)^2$, appears
\cite{Land,Abr}. In other words, despite the interaction between fermions,
the Fermi surface, determined by the ``jump" in the distribution function,
does exist in the system (this is the direct manifestation of the famous Migdal
theorem).

The situation changes drastically at lower dimensions and/or in the case of
a very strong interaction between fermions. As for the low
dimensions, the most convincing example is the 1D conductor, where even
the weakest electron--electron interaction causes the disappearance
of the jump in the distribution function \cite{Luttin}. Such a FL
(called so Luttinger), strictly speaking cannot be referred to
as FL since all  its excitations belong to those of the bose--type
\cite{Luttin,Hald,Volovik}.

The example, demonstrating the influence of an interaction, is the
strongly correlated Hubbard model. On--site repulsion, $U$, in this
model acts on all particles in the system in contrast to the case of a
generally accepted FL where the interaction is actual only in the small
vicinity of the Fermi surface. As to the result,
such a repulsion leads to a considerable
reduction of the QW. When the repulsion increases --- the QW decreases
\cite{Liu,Mart} and in the limiting case $U\to\infty$, the Fermi surface
disappears  completely ($Z=0$) at any fixed density of fermions
\cite{Kuz,Izum}.

As for the HTSCs, according to much experimental data, neither Landau no
Luttinger FL can explicate them. What is interpreted as QW in the case
of HTSCs is relatively small ($Z\simeq 0.4-0.6$ \cite{Ols,Bat}), so the
corresponding FL could not be considered as Landau FL. On the other hand,
it is not Luttinger FL since the QW is not small enough. In addition,
quasiparticles themselves are less stable than in the case of ordinary
Landau FL. Actually, the damping of them in a close vicinity of the Fermi
surface at $T=0$ looks like $\gamma\sim |\epsilon-\epsilon_F|^{\alpha}$, with
$\alpha\simeq 1$ \cite{Bat,Kamp}; this differs from the usual FL power law
with $\alpha=2$. Such a FL with these exceptional characteristics and a
qualitatively different ``intermediate" behavior is now called ``marginal
FL" \cite{Volovik,Varma,Varma1,Ander}.

If one assumes that the marginal FL is relevant to HTSCs, then a
sufficient number of observable physical properties of HTSCs can be quite
naturally explained. Up to the present, however, the latter has been
achieved only in the
framework of some phenomenological theory \cite{Varma}, containing
a lot of additional assumptions, the physical meaning of which has yet to
be clarified.

In what follows we consider the simplest 2D field theoretical model
in the normal phase with a weak inter--fermion attraction (its nature
is not specified here). As will be seen, the Green function of the
order parameter fluctuations (OPF), responsible for the attraction
between fermions in our model, turns out to be consistent with the very
general requirements of the phenomenological theory
\cite{Varma}, being capable of explaining many HTSC properties. Here we
should note, however, that in the cited paper the interaction between fermions
put down due to the exchange by charge or spin fluctuations. Though it is
crucial for the interpretation of some experimental data, as well as for the
calculation of the Green function itself, the latter does not play any
important role. Therefore, below we shall restrict ourselves to studying only
those properties of HTSCs which can be extracted from the Green function of
interacting fermions. In order to be able to consider the other properties,
as in Ref.~\cite{Varma}, we would need to calculate corresponding charge
and spin correlation functions. But that, in fact, is beyond the scope of
this paper.

Below it will be shown that taking into account the interaction mediated
by the OPF leads to a considerable reduction of the QW even at a very weak
attraction between fermions. This, we believe, is the result of
dimensionality lowering, what is accompanied by the enhancement of the
role of any attractive interaction. Investigating the dependence of the QW
on the density of fermions, we shall come to a quite natural conclusion
that decreasing the density leads to decreasing the QW and
increasing the damping. Note that just these two features: (i) a
quasi--2D character and (ii) a relatively low density of carriers, are
as a rule emphasized when properties of  HTSCs are discussed \cite{Phys}.

Lastly, maybe the most interesting result of our paper is a linear
temperature dependence of quasiparticle damping. It is just this feature
that was in fact {\em a priori}
accepted in the phenomenological description \cite{Varma} for explaining the
surprize linear temperature dependence of resistivity observed in high--$T_c$
compounds.

Throughout the paper we use units in which $\hbar=1$ and $k_B=1$.

\section{Model and general discussion}

To solve the problem analytically as far as possible, we choose the
simplest field theoretical model of fermions with attraction. The
corresponding Hamiltonian is \cite{Abr}:
\beq
\hat{H}&=&\int d^2 r {\hat{\cal H}}({\bf r}),\\
{\hat{\cal H}}({\bf r})&=&-\psi^{\dagger}_{\sigma}
\left(\frac{\nabla^2}{2m}+\mu\right)\psi_{\sigma}
-\frac{g}{2}(1-\delta_{\sigma\sigma_1})
\psi^{\dagger}_{\sigma}\psi^{\dagger}_{\sigma_1}\psi_{\sigma_1}\psi_{\sigma},
\label{eq:ham1}
\eeq
Here we use the standard notations: $m$ --- for the effective fermion
mass, $\mu$ --- for the chemical potential, $\sigma=\uparrow$ or $\downarrow$
denotes a spin variable. The interacting constant, $g$, is positive what
corresponds to the attraction between fermions.

Introducing Nambu spinors for fermion field
$\Psi^{\dagger}=(\psi^{\dagger}_{\uparrow},\psi_{\downarrow})$, one can
rewrite Eq.(\ref{eq:ham1}) in a more convenient form:
\beq
{\hat{\cal H}}({\bf r})&=&-\Psi^{\dagger}\tau_{z}
\left(\frac{\nabla^2}{2m}+\mu\right)\Psi
-g\Psi^{\dagger}\tau_{+}\Psi\Psi^{\dagger}\tau_{-}\Psi,
\label{eq:ham2}
\eeq
where $\tau_z$, $\tau_{+}\equiv (\tau_x+i\tau_y)/2$,
$\tau_{-}\equiv (\tau_x-i\tau_y)/2$ are Pauli matrices.

Below we shall use the functional integral approach along with the
Matsubara thermal technique. Thus, the partition function is
expressed through the Hamiltonian as:
\beq
Z(T)=\int[d\Psi^{\dagger}d\Psi]\exp\bigg[-\int\limits_{0}^{\beta}d\tau\int
d^2 r\left(\Psi^{\dagger}\partial_{\tau}\Psi+{\hat{\cal H}}({\bf r})
\right)\bigg],
\qquad \beta\equiv \frac{1}{T},
\label{eq:z(T)}
\eeq
where $T$ is the temperature and $[d\Psi^{\dagger}d\Psi]$ denotes the
measure of the functional integration over the Grassmann variables
$\Psi$ and $\Psi^{\dagger}$, satisfying the antiperiodic
boundary conditions: $\Psi(\tau;{\bf r})=-\Psi(\tau+\beta;{\bf r})$ and
$\Psi^{\dagger}(\tau;{\bf r})=-\Psi^{\dagger}(\tau+\beta;{\bf r})$.

To be able to construct arbitrary Green functions one should add to
(\ref{eq:z(T)}) classical sources associated with the
Grassmann fields in the Hamiltonian (\ref{eq:ham1}).  However, we
shall not write them down explicitly, even though it would
facilitate the understanding of some key formulae below.

Using now an auxiliary Hubbard--Stratonovich complex scalar field
in the usual way, one can represent the Eq.(\ref{eq:z(T)}) in
an equivalent form:
\beq Z(T)&=&\int[d\Psi^{\dagger}d\Psi d\Phi
d\Phi^{*}]\exp\bigg[ -\int\limits_{0}^{\beta}d\tau\int d^2 r
\bigg(\frac{|\Phi|^2}{g}+\nonumber\\
&&+\Psi^{\dagger}\left[\partial_{\tau}-\tau_z
\left(\frac{\nabla^2}{2m}+\mu\right)+\tau_{-}\Phi+\tau_{+}\Phi^{*}
\right]\Psi\bigg)\bigg].                      \label{eq:partF}
\eeq
The main virtue of this representation
is a nonperturbative introduction of the
composite fields: $\Phi=-g\Psi^{\dagger}\tau_{+}\Psi$ and
$\Phi^{*}=-g\Psi^{\dagger}\tau_{+}\Psi$ and possibility to develop
the consistent approach.  Specifically, the expression
(\ref{eq:partF}) turns out to be rather convenient for studying such
a nonperturbative phenomenon as, for example, superconductivity.
In this case a complex Hubbard--Stratonovich field naturally
describes the order parameter arrising due to formation of
pairs (to some extent, the Cooper ones), or composite bosons. The
average value of $|\Phi|$ is proportional to the density of
these bosons, on one hand, and determines the gap in the
one--particle fermi--spectrum, on the other.  In the normal phase
the homogeneous order parameter disappears, {\em i.e.}  $|\Phi|$
equals zero, but the OPF inevitably take place even there.
Intuition, however, suggests  that these fluctuations should play the
crucial role in a small neighbourhood of the corresponding phase
transition. As it will be shown, this is in fact the case in our model.
Moreover, OPF not only exist but also have an influence on different
characteristics of normal phase on the whole.

Before going further, it may help to recall the most essential
property of 2D systems.  The question is the famous
Mermin--Wagner--Hohenberg (MWH) theorem which forbids the appearance
of a complex (in our case --- charged) order parameter $\Phi$ at any
finite temperature \cite{Merm,Hohen}. If we studied the problem in
the mean field approximation we would obtain a finite critical
temperature $T_c$, for instance, of the superconducting phase
transition. Taking into account the next to leading order
approximation immediately reduces $T_c$ to zero.  Such a
manifestation of the MWH theorem may make the study of pure 2D superconductors
at $T\neq 0$ meaningless. However, as was shown recently,
the situation here is much more subtle. Actually, 2D metals undergo
two phase transitions \cite{MacK}.  The first one is connected with violating
only some discrete symmetry and the appearance of  a neutral order parameter
(modulus of the Hubbard--Stratonovich field, $|\Phi|$ ). It is this
transition that was usually interpreted in the mean field
approximation as the appearance of superconductivity. Despite this
modulus being different from zero, the field $\Phi$ on the average
is zero, due to the temporal and spatial fluctuations of its
phase factor. We would call such a state of the system by the term
``molecular gas", emphasizing by this notion that there is no phase
coherence in the ``Cooper pair" dynamics\footnote{It must be noted
that in Ref.~\cite{Triv} there was introduced the temperature $T_p$
which characterized the energy scale of composite boson formation.}.
The second phase transition is connected with the appearance of the
so called Berezinsky--Kosterlitz--Thouless (BKT) phase \cite{Berez,Kost}.
Developing of such a phase is just accompanied also by the
superconductivity \cite{MacK}.

In what follows we shall study only the normal phase of 2D metal,
{\em i.e.} the phase where even the neutral homogeneous order
parameter is absent.  We believe that precisely this phase of pure 2D
model represents the true  normal phase of quasi--2D HTSCs more
adequately.

Below we shall use a generally accepted assumption that the approximation
taking into account only the quadratic (Gauss) fluctuations of $\Phi$ (around
zero in our normal phase) describes the system quite well. Thus,
integrating out the fermion fields in (\ref{eq:partF}), one obtains:
\beq
Z(T)&=&\exp(-S_{0})\int[d\tilde{\Phi} d\tilde{\Phi}^{*}]\exp\bigg[
-\int\limits_{0}^{\beta}d\tau_1\int\limits_{0}^{\beta}d\tau_2\nonumber\\
&&\times\int d^2 r\int d^2 r
\tilde{\Phi}^{*}(\tau_1;{\bf r}_1)
\Gamma^{-1}(\tau_1-\tau_2;{\bf r}_1-{\bf r}_2)\tilde{\Phi}(\tau_2;{\bf r}_2)
\bigg],                   \label{eq:PFquad}
\eeq
where
\xq
S_0=Tr Ln[G_0]    \label{eq:so}
\zq
is the one--loop effective action, and
\beq \Gamma^{-1}(\tau;{\bf
r})=\frac{1}{g}\delta(\tau)\delta({\bf r}) +tr\bigg[G_0(\tau;{\bf
r})\tau_{-}G_0(-\tau;-{\bf r})\tau_{+}\bigg]  \label{eq:Gam0}
\eeq
is the inverse Green function (propagator) for OPF.
Both quantities (\ref{eq:so}) and (\ref{eq:Gam0}) are expressed through
the free fermion Green function which obeys the equation:
\xq
\bigg[-\partial_{\tau}+\tau_z\left(\frac{\nabla^2}{2m}+\mu\right)\bigg]
G_0(\tau;{\bf r})=\delta(\tau)\delta({\bf r})
\label{eq:G0}
\zq
with the following boundary condition:
\xq
G_0(\tau+\beta;{\bf r})=-G_0(\tau;{\bf r}).
\label{eq:effpart}
\zq
The integration in (\ref{eq:PFquad}) can be formally performed. The result
is
\xq
Z(T)=\exp(-S_{0}-Tr Ln[\Gamma^{-1}]).
\zq
If we could have obtained the explicit expression for the partition function
we could, in principle, obtain all thermodynamical functions but in
reality it is usually impossible to do this. Therefore, we shall study
more simple objects, namely, the Green functions which can also give
the information about the system.

To avoid possible further misunderstanding, we shall write down the formulae
which are used throughout the paper for Fourier transformations, connecting
coordinate and momentum representations:
\beq
F(i\omega_n,{\bf k})&=&\int\limits_{0}^{\beta}d\tau\int d^2 r
F(\tau,{\bf r})\exp(i\omega_n\tau-i{\bf k}{\bf r}),
\label{eq:fomeg}\\
F(\tau,{\bf r})&=&T\sum_{n=-\infty}^{\infty}\int\frac{d^2 k}{(2\pi)^2}
F(i\omega_n,{\bf k})\exp(-i\omega_n\tau+i{\bf k}{\bf r}),
\eeq
where $\omega_n=\pi T(2n+1)$ are fermion (odd) Matsubara frequencies. In
the case of bosons they should be replaced by even ones: $\Omega_n=2\pi Tn$.

\section{Green function of OPF}

The Green function for free fermions in the momentum representation
(\ref{eq:fomeg}) takes, as follows from (\ref{eq:G0}), the very simple form:
\xq
G_0(i\omega_n,{\bf k})=\frac{1}{i\omega_n-\tau_z\left(\frac{{\bf k}^2}{2m}
-\mu\right)}.
\label{eq:green0}
\zq
Substitution of this expression into (\ref{eq:Gam0}) leads to the
explicit formula for the inverse Green function for OPF in the
momentum representation:
\beq
\Gamma^{-1}(i\Omega_n;{\bf K})=\frac{1}{g}-\frac{1}{8\pi^2}\int d^2k
\frac{\tanh(\beta\xi_{+}/2)+\tanh(\beta\xi_{-}/2)}{\xi_{+}+\xi_{-}-i\Omega_n}
\label{eq:GamP}
\eeq
where $\xi_{\pm}\equiv \frac{1}{2m}({\bf k}\pm{\bf K}/2)^2-\mu$;
${\bf k}$ and ${\bf K}$ are relative and total momenta of a pair,
respectively.  The divergence at large momenta, ${\bf k}$, in
(\ref{eq:GamP}) is regularized by the ultraviolet cutoff, $W$, which
may be considered as the free fermion band width. Assuming that the
cutoff is much bigger than all other characteristic scales of the
problem, we can get rid of it after coupling
constant renormalization and next turning $W$ to infinity (see,
for example, \cite{Gorbar}). So, expressing the coupling constant in
the form
\xq
\frac{1}{g}=\frac{m}{4\pi}\ln\frac{2W}{|\epsilon_b|}, \label{eq:eb}
\zq
where $\epsilon_b$ is the two--particle bound state energy,  and
substituting it into Eq.(\ref{eq:GamP}), one can immediately take the
limit $W\to\infty$. The  detailed calculation can be found in Appendix A
(formula (\ref{eq:Gtotal})). As is easily seen, the expression obtained is
rather complicated for use or even for qualitative analysis.  However, there
is one particular aspect of it which should be emphasized here --- the
asymptotics of the imaginary part of $\Gamma^{-1}(\Omega;{\bf K})$.
The explicit expression for it is:
\beq Im\Gamma^{-1}(\Omega;{\bf K})&=&-\frac{m}{4}\theta(\Omega+2\mu-{\bf
K}^2/4m)
\tanh(\beta\Omega/4)\times\nonumber\\ &&\times[1-f(\Omega+2\mu-{\bf
K}^2/4m,K)], \label{eq:Impart}
\eeq
where the function $f(x,y)$ (see Appendix A)
varies from zero to one and vanishes as $x,y\to 0$ or $x,y\to
\infty$. It is easy to be convinced that the asymptotics of
(\ref{eq:Impart}) at $\Omega\to 0$ and $\Omega\to \infty$ coincide
exactly with those postulated in Ref.~\cite{Varma}. Recall
that the phenomenological theory which was proposed in that paper
explains in principle a lot of experimental data on HTSCs.  Therefore,
one can hope that our microscopic approach will also be able to
describe correctly some observable properties of HTSCs.

In order to calculate the Green function of interacting fermions (see the
next section) one still needs to find some rather simple approximation for
OPF Green function. The most commonly used one is the so called the
derivative expansion, which is valid at small frequencies and momenta.
To obtain a consistent expression we need first to construct the
retarded real--time Green function (by means of
analytical continuation $i\Omega_n\to\Omega+i0$) and then to
expand it at ${\bf K}\to 0$ and $\Omega\to 0$. Using the explicit
formula (\ref{eq:Gtotal}), we arrive at the following representation:
\xq
\Gamma^{-1}(\Omega;{\bf K})=a+b\frac{{\bf K}^2}{4m}-c\Omega,
\label{eq:DerEx} \zq where \beq
a&=&\frac{m}{4\pi}\left[\ln\frac{\pi}{\beta|\epsilon_b|\gamma}-
\int\limits_{0}^{1}\frac{dx}{x}\tanh\frac{x\beta\mu}{2}\right],
\label{eq:asmall}\\
b&=&\frac{m}{8\pi\mu}\left[\tanh\frac{\beta\mu}{2}+\frac{\beta\mu}{2}
+\frac{(\beta\mu)^2}{4}\int\limits_{-\beta\mu/2}^{\infty}\frac{dx}{x^2}
\tanh^2x\right],     \\
c&=&\frac{m}{8\pi\mu}\left[\int\limits_{1}^{\infty}\frac{dx}{x^2}
\tanh\frac{x\beta\mu}{2}+i\frac{\pi\beta\mu}{2}\theta(2\mu)\right],
\label{eq:coefc}
\eeq
and $\ln\gamma\simeq 0.577$ is the Euler constant. Note
that here and henceforth the same symbols are used for both the
real--time and thermal Green functions.

The remarkable feature of the expansion (\ref{eq:DerEx})
is the existence of a term that is
linear in frequency (the next, quadratic, term is omitted in
(\ref{eq:DerEx})). We stress this point here because sometimes the
papers devoted to nonrelativistic systems appear (for example,
\cite{Stoof}) where even the possibility of such a term in derivative
expansion (or, what is the same, in the composite boson effective action) is
not
assumed. Perhaps this is the result of the uncritical use of some relativistic
methods in nonrelativistic systems. In any case, there is no any reason for
omitting this term in the model under consideration.  Note that the importance
of such a term has been recently emphasized also in Ref.~\cite{Randeria,Stone},
where a similar model for  $3D$ metal was investigated. Finally, the appearance
of a term that is linear in frequency seems to be quite natural in
Galilean--invariant models, and its absence should be regarded as being
suspicious, if there is no any rather convincing reason for its absence.

We note here that the
condition, indicating the appearance of instability in the system, reads
as $a(\mu,T)=0$, {\em i.e.} (see (\ref{eq:asmall})) in our
case\footnote{In Ref.~\cite{Gorbar} just such an equation
was considered as one of the system of self--consistent equations
for determining the values of $T_c$ and $\mu_c\equiv\mu(T_c)$ in
mean--field approximation at constant density of particles.}:
\xq
\ln\frac{\pi}{\beta|\epsilon_b|\gamma}-
\int\limits_{0}^{1}\frac{dx}{x}\tanh\frac{x\beta\mu}{2}=0.
\label{eq:stabil}
\zq
This equality defines a curve (below it will be called ``the
stability line") which separates two different regions on the
$\mu$--$T$ plane (see Fig.1).  The region ``N" in Fig.~1 corresponds
to the normal phase of the system where there is no order parameter.
Inside the second region, denoted by ``OP", some nonzero order
parameter appears. It has to be noted that the developing of the
BKT phase also takes place somewhere inside the OP  area.

As for the explicit dependence $\mu=\mu(T)$ on the stability line
determined by the equation (\ref{eq:stabil}), we can only give
asymptotics at both low and high temperatures:
\beq
\mu\simeq-\frac{|\epsilon_b|}{2}\qquad\mbox{at}\qquad T\ll
|\epsilon_b|,\\
\mu\simeq\frac{\pi^2T^2}{2\gamma^2|\epsilon_b|}\qquad\mbox{at}\qquad
T\gg |\epsilon_b|.
\eeq
To understand the physical meaning of the
stability line one needs to depict the curve which
describes the physical dependence of the chemical potential (so far
it was treated as an independent external parameter) on the
temperature at fixed density of carriers in our system. The equation
that determines this second curve is \xq
n_f(\mu,T)=\frac{T}{(2\pi)^2}\sum_{n=-\infty}^{\infty}\int d^2p~
tr[\tau_z G(i\omega_n,{\bf p})],\label{eq:SecondCurve}
\zq
where $G(i\omega_n,{\bf p})$ is the interacting fermion Green
function expressed in its turn through the temperature and the
chemical potential. Note that, after taking into account
the expression (\ref{eq:effpart}) and the definition for the bare
density: $n_f=T/V\partial\ln{Z(T)}/\partial\mu$,
Eq.~(\ref{eq:SecondCurve}) can also be represented as the sum of two
terms (see also \cite{Randeria,GLS}). The first term, $n_F$, is expressed
through the free fermion Green function and the second one, $n_B$, ---
through
the Green function of OPF (or composite bosons). Thus, one may interpret
such a situation as if a dynamical balance
$n_f(\mu,T)=n_F(\mu,T)+2n_B(\mu,T)$ in the distribution of the
initial noninteracting particles over fermion and boson degrees of
freedom is established in the system.

As the first approximation, we could substitute the free
fermion Green function (\ref{eq:green0}) into (\ref{eq:SecondCurve}) what
is equivalent to omitting the bosonic term in the expression for the
distinctive ``conservation law"
$n_f=n_F+2n_B$.  As a result, one can see that the curve which
describes the dependence of the chemical potential on the temperature looks
somewhat similar to the dashed line in Fig.~1.  Increasing or
decreasing the density of carriers in the system would lead to moving this
curve up or down, respectively.

Turning back to the physical meaning of the stability line, one can
be convinced that, changing the temperature in the system corresponds
to removal to the right along the dashed line in Fig.~1. Thus, we
can naturally conclude that our system is in its normal phase at
temperatures that are high enough. On the contrary, decreasing the
temperature corresponds to displacement to the left along the dashed
line.  When the system proves to be in  the point where this
line intersects the stability line a neutral order parameter
appears. Subsequent removal, in OP region, will result in the
transition into BKT phase (the second stability line, which must be
calculated from another equation \cite{MacK}, is not depicted in
Fig.~1).

\section{Fermion Green function}

The general form for the inverse thermal Green function of
interacting fermions reads (compare with (\ref{eq:green0})):
\xq
G^{-1}(i\omega_n,{\bf p})=i\omega_n-\tau_z\left[\frac{{\bf p}^2}{2m}
-\mu+\Sigma(i\omega_n,{\bf p})\right], \label{eq:ferGr}
\zq
where the self--energy operator for interacting fermions,
$\Sigma(i\omega_n,{\bf p})$, can be already expressed through the
free Green function of fermions and OPF Green function by means of
the following formula:  \beq \Sigma(i\omega_n,{\bf
p})&=&\frac{\tau_z}{\beta}\sum_{n'=-\infty}^{\infty}
\int\frac{d^2k}{(2\pi)^2}\bigg[\tau_{+}G_0(i\omega_{n'},{\bf
k})\tau_{-} \Gamma(i\Omega_{n-n'},{\bf p}-{\bf k})+\nonumber\\
&&+\tau_{-}G_0(i\omega_{n'},{\bf k})\tau_{+}
\Gamma(i\Omega_{n'-n},{\bf k}-{\bf p})\bigg].
\label{eq:self}
\eeq
To trace the complete derivation of this expression one
needs to use the standard Feynman technique for the system
given by (\ref{eq:partF}) \cite{Fradkin}. Now the substitution of the
OPF Green function in its derivative approximation (\ref{eq:DerEx})
into the last formula leads to the integral representation for the
self--energy operator (\ref{eq:self}):
\beq \Sigma(i\omega_n,{\bf
p})=\frac{1}{2} \int\frac{d^2k}{(2\pi)^2}\frac{\coth\frac{\beta}{2c}
\left(a+b\frac{({\bf p}-{\bf k})^2}{4m}\right)-\tanh\frac{\beta}{2}
\left(\frac{{\bf k}^2}{2m}-\mu\right)}
{a+b\frac{({\bf p}-{\bf k})^2}{4m}-c\left(\frac{{\bf k}^2}{2m}-\mu\right)
-ic\tau_z\omega_n}.
\label{eq:regG0}
\eeq
where the summation over the Matsubara frequencies was performed.
Assuming that the result of integration in the last expression does not
strongly depend on the particular form of the smooth function in the
numerator, we can obtain the following approximate self--energy operator:
\beq
\Sigma(i\omega_n,{\bf p})=\int\frac{d^2k}{(2\pi)^2}
\frac{\theta(\mu-\frac{{\bf k}^2}{2m})}
{a+b\frac{({\bf p}-{\bf k})^2}{4m}-c\left(\frac{{\bf k}^2}{2m}-\mu\right)
-ic\tau_z\omega_n}.
\label{eq:regG1}
\eeq
To arrive at this expression, we have carried out some formal limit
of zero temperature in (\ref{eq:regG0}). In other words, only
the explicit dependence of this expression on the temperature through
the arguments of the hyperbolic functions was subjected to this
limit: $\coth\beta(\ldots)\to 1$,  $\tanh\beta(\ldots)\to
sign(\ldots)$. Of course, such a treatment introduces some
uncertainty in the quantitative analysis. But the advantage of
making further step by means of analytical manipulations is
so attractive that we would like to expect the reliability of such
an approximation.  After simplification mentioned, the integration
over momenta in Eq.(\ref{eq:regG1}) can be performed explicitly and
the result is written down in Appendix B (formula (\ref{eq:GreenInter})).

The real--time retarded fermion Green function in a close vicinity
of its pole (as follows from (\ref{eq:ferGr})) has the standard form:
\xq
G(\omega,{\bf p})=\frac{Z}{\omega-\tau_z\left(E_p-i\gamma_p\right)},
\label{eq:thermG}
\zq
where $Z$ is the introduced above QW, $\gamma_p$ is the quasiparticle
damping and $E_p$ is some function determining the low--energy
one--particle spectrum. All these functions in (\ref{eq:thermG}) are
expressed through the self--energy operator (\ref{eq:regG1}):
\beq
Z^{-1}&=&1-\tau_zRe\left.\frac{\partial
\Sigma(\omega;p_0)} {\partial \omega}
\right|_{\omega=0},\label{eq:Zp}\\ E_p&=&Z\left(1+2mRe\frac{\partial
\Sigma(0;p_0)}{\partial p_0^2}\right)
\left(\frac{p^2-p_0^2}{2m}\right),\label{eq:Ep}\\
\gamma_p&=&-Z Im\left(\Sigma(0;p_0)+\frac{\partial \Sigma(0;p_0)}
{\partial p_0^2}(p^2-p_0^2)\right),
\label{eq:gammap}
\eeq
where $p_0$ is renormalized ``Fermi" momentum, which is the solution to
the equation $E_p=0$, or
\xq
\frac{p_0^2}{2m}=\mu-Re\Sigma(0;p_0).
\label{eq:p0}
\zq
Note that this equation could be solved, in principle, at any chosen
values of the chemical potential and the temperature. However, taking
into account physical reasons, we restrict ourselves to solving the
equation (\ref{eq:p0}) for two sets of parameters $\{T,\mu\}$ only.
The first one corresponds to those values of the temperature and the chemical
potential which lie on the stability line. From the physical point of
view, this choice implies that the system is kept close to the
critical region, but the density of carriers in the system increases
monotonically at moving up along the stability line, or decreases in
the opposite case. As a result, one will obtain the qualitative
dependence of the near--critical properties of 2D metal with
inter--fermion attraction on the density of carriers.  The second set
relates to the curve (the dashed line in Fig.~1), where the density of
carriers is constant. This gives us the information about the dependence
of 2D metal properties on the temperature.

\section{Numerical results and discussion}

We start from the analysis of near--critical properties at different
carrier densities. As was noted above (see Section~3), the stability
line is determined by the condition $a(\mu,T)=0$. It is easy to check
that being taken into account this condition simplifies the analysis
essentially. Indeed, the only quantities which one needs to
know for the calculation of $Z$, $E_p$ and $\gamma_{p_0}$ (defined in
(\ref{eq:thermG})) are the self--energy operator and its first
derivative with respect to frequency at $\omega=0$
(see (\ref{eq:Zp})--(\ref{eq:gammap}) ). Both these
quantities are given in Appendix~B. Performing the analysis,
we come to the conclusion that in the case of parameters belonging
to the stability line the relevant self--energy operator does not
depend on the momentum and, as a result, the equation (\ref{eq:p0})
transforms into expression for the momentum $p_0$. Substituting then
this momentum into (\ref{eq:Zp}), (\ref{eq:Ep}) and (\ref{eq:gammap})
we would obtain complete information about the Green function for
the interacting fermions. We, however, shall not write down
corresponding cumbersome
complex expressions, but instead present the graphs
for $\tilde{\xi}=\left({\bf p_0}^2/2m-\mu\right)/\mu$, $Z$ and
$\gamma_{p_0}$ as functions of $\beta\mu$ (Fig.~2--Fig.~4, respectively).

Note that only the region of positive values of
the chemical potential is considered. Our investigation fails in the region
near the point $\mu=0$, since the derivative expansion (\ref{eq:DerEx})
is not valid there. Though there are no obstacles for studying the region
of negative values of the chemical potential separately, we shall not
do that because we are interested in obtaining some qualitative
information about the properties of the system proceeding just from
the general monotonic dependence of the chemical potential on the
particle density on the stability line.  The singularity in
Eq.(\ref{eq:DerEx}) at $\mu=0$ (because of $c\to\infty$, see
(\ref{eq:coefc})) forbids us to go into the region of negative
chemical potential continuously.

Since the parameter $\beta\mu$
monotonically changes along the stability line, the graphs in
Fig.~2--Fig.~4 can be thought of as a qualitative dependence of corresponding
quantities on the density of carriers in the system. Thus, we conclude that
when the carrier density is high the system naturally tends to become
conventional FL with a rather small damping of the particles and QW
practically equals one.  In the opposite case, damping increases and
QW reduces, {\em i.e.} marginal features of FL really appear.

 Let us now turn to studying the properties of the system at
fixed density of carriers. This case turns out to be much more
difficult than the previous one. In particular, the equation
(\ref{eq:p0}) cannot be solved so easily as before. In fact, one
should consider the system of two self--consistent equations
(\ref{eq:SecondCurve}) and (\ref{eq:p0}) at each chosen temperature.
To simplify the system, one can use a quite natural
approximation in which the additional equation
(\ref{eq:SecondCurve}) is taken in the mean field form, {\em i.e.}
instead of interacting fermion Green function we shall use the free one.
But even in this case the equation (\ref{eq:p0}) cannot be solved
explicitly. Solving it numerically and substituting into (\ref{eq:Zp})
and (\ref{eq:gammap}), we obtained the dependence of
$\tilde{\xi}=\left({\bf p_0}^2/2m-\mu\right)/\mu$, $Z$ and $\gamma_{p_0}$
on the temperature. The corresponding graphs are represented in
Fig.~5--Fig.~7 for several different densities: (1)
$\epsilon_F/|\epsilon_b|=500$; (2) $\epsilon_F/|\epsilon_b|=750$; (3)
$\epsilon_F/|\epsilon_b|=1000$, where $\epsilon_F\equiv\pi n/m$ by
definition. All curves start at the corresponding critical
temperatures. For example, it is seen that in the wide range of $T$'s
the QW considerably differs from 1, essentially depending on $T$.

But the most amazing, to our mind, result of the numerical analysis is a linear
temperature dependence of the quasiparticle damping within a rather
wide range of temperatures (from $T_c$ up to about $4T_c$ and even
higher). As was noted above, it is just such a linear dependence that
was supposed in Ref.~\cite{Varma} for explaining the unusual
temperature dependence of resistivity of HTSCs\footnote{This result does not
seem to correspond to the marginal Fermi liquid definition (see
Introduction). It must be kept in mind, however, that the definition mentioned
refers
to the case $T=0$ which should be excluded from our consideration by physical
reasons. As to $T\neq 0$ the quasiparticle damping proves to be nonzero
(see (\ref{eq:gammap})) even at Green function pole $p=p_0$.}.
Our calculations show also that the decreasing of the ratio
$\epsilon_F/|\epsilon_b|$ rather quickly breaks the linear dependence of
damping $\gamma_{p_0}$ on $T$ what can be probably explained by the sharp
decreasing $\mu$ up to so small values where the derivative expansion fails.

Of course, numerical results presented here cannot, by themselves, give us a
completely convincing answer concerning the physical reasons for an
explanation of the peculiarities of the 2D metal model. Our hypothesis
is quite simple and seems to be rather natural: the most important features
which are responsible for the marginal FL behavior
are lowered dimensionality and, of course, attractive character of
inter--particle interaction. Indeed, the two--dimensionality has
resulted in the asymptotics of the imaginary part  of the inverse
Green function (\ref{eq:Impart}):
\beq
Im\Gamma^{-1}(\Omega;{\bf K})&\simeq&-\frac{m\beta\Omega}{16},
\qquad\mbox{at}\qquad\Omega\to 0\\
Im\Gamma^{-1}(\Omega;{\bf K})&\simeq&-\frac{m}{4},\qquad\mbox{at}
\qquad\Omega\to \infty,
\eeq
which were required in Ref.\cite{Varma} (pay attantion that the Green
functions for spin and charge fluctuations  were used  in the paper
cited). As to the attraction, it may play the important role even in
3D models but in 2D case its effect proves to be decisive. While in
the former case OPF are very important mainly in the superconducting
phase, in the latter one they turn out to be of great importance in
the normal phase too.

\section{Conclusion}

In this paper we studied a model of a 2D metal with a local weak
attraction between fermions  in the normal phase. Within the
framework of rather natural approximations, we have shown that such a
system displays properties which differ essentially (mainly in the
vicinity of the stability line) from the conventional properties of
Landau FL. In particular, the QW noticeably differs from
1 and the damping of quasiparticles is rather large.  The
manifestation of this effect becomes more essential as the density
of particles decreases.

We also investigated the temperature dependence of the QW and the
quasiparticle damping. The most important result, we believe, is the behavior
of obtained curves for the quasiparticle damping, exhibiting features which
can explain the linear temperature dependence of resistivity which is
observed at all temperatures of HTCS normal phase existance.

Since through the paper we restricted our consideration of the 2D
model with non--retarded inter--particle interaction only which, as
is well known, is valid in the case of rather small Fermi energy,
it seems to be very interesting and important to investigate the
influence of an indirect interaction between fermions (mediated, for
example, by the phonons, magnons or some other bosons and resulted
in a retarded attraction) and its possible connection with the OPF in
the normal phase too. One more rather urgent for HTCSs problem is to
study the possible effects when the OPF have non--trivial (for example,
$d$--wave) symmetry. At last, we would like also to note that since
our approach involves some intuitive assumptions (even though they
are commonly used in solid state physics), more consistent
consideration would be useful.

One of us (V.P.G.) would like to acknowledge Prof.~D.Atkinson for
critical reading the manuscript.
The research was supported in part by the Grant No.K5O100 of the
International Science Foundation and by the International Soros Science
Education Program (ISSEP) through the Grant No.PSU052143.

\section*{Appendix A}

Here we derive the renormalized OPF Green function. Substituting  the
coupling constant (\ref{eq:eb}) into (\ref{eq:GamP}), one obtains
\beq
\Gamma^{-1}(i\Omega_n;{\bf K})&=&\frac{m}{4\pi}\bigg[
\ln\frac{2W}{|\epsilon_b|}-\int\limits^{2W}_{0} dx
\frac{\tanh\left(\beta(x+\frac{{\bf K}^2}{4m}-2\mu)/4\right)}
{x+\frac{{\bf K}^2}{4m}-2\mu-i\Omega_n}+
\nonumber\\
&&+\int\limits_{0}^{2W} dx\frac{
\tanh\left(\beta(x+\frac{{\bf K}^2}{4m}-2\mu)/4\right)}
{x+\frac{{\bf K}^2}{4m}-2\mu-i\Omega_n}
f(x,K)\bigg], \label{eq:regG}
\eeq
where the function $f(x,K)$ is
\beq
f(x,y)=\frac{2}{\pi}\int\limits_{0}^{\pi/2}d\phi
\frac{\cosh\left(\frac{y\beta\sqrt{x}}{2\sqrt{m}}\cos\phi\right)-1}
{\cosh\left(\beta(x+\frac{{\bf p}^2}{4m}-2\mu)/2\right)+
\cosh\left(\frac{y\beta\sqrt{x}}{2\sqrt{m}}\cos\phi\right)}.
\eeq
 (see also (\ref{eq:Impart})).
As was pointed above and can be easily checked directly this
function varies from zero to one and vanishes when $x$ or $y\to 0$ as
well as when $x$ or $y\to \infty$.

Using identities:
\beq
\frac{1}{x+\frac{{\bf K}^2}{4m}-2\mu-i\Omega_n}&=&
\frac{1}{x+\frac{{\bf K}^2}{4m}-2\mu}+\nonumber\\
&&+\frac{i\Omega_n}{\left(x+\frac{{\bf K}^2}{4m}-2\mu\right)
\left(x+\frac{{\bf K}^2}{4m}-2\mu-i\Omega_n\right)},
\eeq
and
\beq
\int\limits_{0}^{\alpha}\frac{dx}{x}\tanh{x}=\ln\frac{4\alpha\gamma}{\pi}-
\int\limits_{\alpha}^{\infty}\frac{dx}{x}(\tanh{x}-1),\qquad \mbox{at}
\qquad\alpha>0,
\eeq
(as earlier $\ln\gamma$ is the Euler constant), one can
take the limit $W\to\infty$ in (\ref{eq:regG}) and come to the final
expression:
\beq \Gamma^{-1}(i\Omega_n;{\bf K})&=&\frac{m}{4\pi}\bigg[
\ln\frac{\pi}{\beta|\epsilon_b|\gamma}+\int\limits^{1}_{0}
\frac{dx}{x}\tanh\left(x\beta(\frac{{\bf K}^2}{4m}-2\mu)/4\right)-
\nonumber\\
&&-i\Omega_n\int\limits_{0}^{\infty}
\frac{dx\tanh\left(\beta(x+\frac{{\bf K}^2}{4m}-2\mu)/4\right)}
{\left(x+\frac{{\bf K}^2}{4m}-2\mu\right)
\left(x+\frac{{\bf K}^2}{4m}-2\mu-i\Omega_n\right)}
+\nonumber\\
&&+\int\limits_{0}^{\infty}
\frac{dx\tanh\left(\beta(x+\frac{{\bf K}^2}{4m}-2\mu)/4\right)}
{x+\frac{{\bf K}^2}{4m}-2\mu-i\Omega_n}f(x,K)\bigg],
\label{eq:Gtotal}
\eeq

\section*{Appendix B}

To calculate the integral (\ref{eq:regG1}), one needs to use the following
table integrals (see, for example, \cite{Ryz}):
\xq
\int\limits_{0}^{\pi}\frac{d\phi}{b+a\cos\phi}=\frac{\pi}{\sqrt{b^2-a^2}}
\qquad \mbox{at} \qquad |b|>|a|,
\zq
and
\xq
\int\frac{dx}{\sqrt{R(x)}}=\frac{1}{\sqrt{c}}\ln\bigg(2\sqrt{cR(x)}+2cx+b\bigg),
\zq
where $R(x)=a+bx+cx^2$ and $c>0$.

Thus, straightforward manipulations lead to the formal result for the
self--energy operator expressed through the elementary functions:
\beq
&&\tilde{\Sigma}(\tilde{\omega};\tilde{\xi})=\frac{4}{\tilde{b}-\tilde{c}}\times\nonumber\\
&&\times\ln\frac{\sqrt{
\frac{(\tilde{b}\tilde{\xi})^2}{4}+(\tilde{a}-\tilde{c}\tilde{\omega})^2+
\tilde{b}(\tilde{\xi}+2)(\tilde{a}-\tilde{c}\tilde{\omega})}
-\frac{\tilde{b}\tilde{c}(\tilde{\xi}+2)+\tilde{\xi}\tilde{b}^2}{2(\tilde{b}-\tilde{c})}
+\tilde{a}-\tilde{\omega}\tilde{c}}
{2(\tilde{a}-\tilde{c}\tilde{\omega})
-\frac{\tilde{c}(\tilde{\xi}\tilde{b}+\tilde{c})}{\tilde{b}-\tilde{c}}},
\label{eq:GreenInter}
\eeq
where the dimensionless functions
\beq
\tilde{a}(\beta|\epsilon_b|,\beta\mu)=\frac{4\pi}{m}a , \qquad
\tilde{b}(\beta\mu)=\frac{4\pi\mu}{m}b ,  \qquad
\tilde{c}(\beta\mu)=\frac{4\pi\mu}{m}c ,\\
\tilde{\Sigma}=\frac{1}{\mu}\Sigma ,\qquad
\tilde{\xi}=\frac{1}{\mu}\left(\frac{{\bf p}^2}{2m}-\mu\right), \qquad
\tilde{\omega}=\frac{\tau_z}{2\mu}\omega,
\eeq
more convenient for numerical calculations, were introduced.
Note that one of the parameters ($\tilde{c}$) in (\ref{eq:GreenInter}) is
complex, so there is some uncertainty in the expression. To avoid it one
needs to clarify which branch of the complex function is chosen. However,
in the main text of the paper only the self--energy
operator and its derivative with respect to frequency at $\omega=0$ and
large value $\beta\mu$ are used. Both required functions at those conditions
are free of the mentioned uncertainty, so we shall not concern about the
difficulty of choosing the needed branch.

When parameters of the system are chosen so that they obey the equation
$\tilde{a}(\beta|\epsilon_b|,\beta\mu)=0$, which identically coincides
with (\ref{eq:stabil}) and determines the stability line, the
self--energy operator and its derivative at $\omega=0$ become very
simple:  \beq
\tilde{\Sigma}(0;\tilde{\xi})=\frac{4}{\tilde{b}-\tilde{c}}
\ln\frac{\tilde{b}}{\tilde{c}}\qquad\mbox{at}\qquad\tilde{\xi}<0;
\eeq
\beq
\left.\frac{\partial\tilde{\Sigma}(\omega;\tilde{\xi})}{\partial\tau_z\omega}
\right|_{\omega=0}=-\frac{4\tilde{c}}{\tilde{\xi}\tilde{b}(\tilde{\xi}+\tilde{c})}-
\frac{4\tilde{b}}{(\tilde{b}-\tilde{c})(\tilde{\xi}\tilde{b}+\tilde{c})}
\qquad\mbox{at}\qquad\tilde{\xi}<0.
\eeq
The expressions for $\tilde{\xi}>0$ can also be obtained from
(\ref{eq:GreenInter}). We do not write down them since they were
not used in our analysis in the main part of the paper because the
equation $\tilde{\xi}=-\tilde{\Sigma}$ (see \ref {eq:p0}) only has a
solution with negative sign.

\end{document}